\documentclass[letter]{aa} 
\usepackage{txfonts}
\usepackage{graphicx,lipsum}
\usepackage[percent]{overpic}
\usepackage{amssymb,amsmath}
\usepackage[version=4]{mhchem}
\usepackage{newtxtext}
\usepackage[varvw]{newtxmath}
\usepackage{xcolor}
\usepackage{caption,subcaption}

\newcommand{\ignore}[1]{}

\def\aap {{A\&A}}

\def\apjl {{ApJL}}

\def\apjs {{ApJS}}

\begin{document}

   \title{ Searching for the grand tack in exoplanetary data }
   \titlerunning{ Searching for the grand tack in exoplanetary data }


   \author{A.J. Cridland\inst{1}\thanks{A.Cridland@lmu.de}
   }
    \authorrunning{ A.J. Cridland }

   \institute{
   $^1$Universit\"ats-Sternwarte M\"unchen, Ludwig-Maximilians-Universit\"at, Scheinerstr. 1, 81679 M\"unchen, Deutschland
                }

   \date{Received \today}


  \abstract
  {
The grand tack model, more generally called the Masset and Snellgrove mechanism, is a planetary migration model whereby two giant planets via interactions with their natal disk migrated to larger orbital radii. While its relevance in our own Solar System remains in question, the fact that the Masset and Snellgrove mechanism is a general hydrodynamical effect implies that it may have occurred in another planetary system. In this study I searched through exoplanet data for evidence of the Masset and Snellgrove mechanism, which requires that (1) the inner of the two planets is more massive than the outer planet; (2) the planets are sufficiently massive that their gravity-induced gap overlaps; and (3) they orbit at sufficiently close radii that their co-rotation regions also overlap. The last two requirements are met when the planets orbit with a 3:2 mean motion resonance. I do not find conclusive evidence for a grand tack-like system, but find some evidence for planet formation at the edge of a planet-induced protoplanetary disk gap in three systems.
}


    \maketitle
%

\section{Introduction}\label{sec:intro}

A natural consequence of the gravitational interaction between young planets and the gas in their natal protoplanetary disks is a reduction in their orbital period (semi-major axis), commonly called planetary migration. The grand tack scenario of the Solar System posits that during the early stage of the Solar System's formation, Jupiter migrated to a smaller orbital radius than its current orbital radius of 5.2 astronomical units (AU; with a period of 11.86 years), thereby impacting the formation of the terrestrial planets, including the low mass of Mars \citep{Walsh2011}. The migration trajectory of Jupiter was then reversed by the young, forming Saturn as its mass grew and its migration brought it into 3:2 mean motion resonance (MMR). At this particular MMR, and with a sufficiently massive Saturn, the pair migrated together outwards to the larger orbital periods we see today. 

The grand tack scenario is a specific example of what is known as the Masset and Snellgrove mechanism, as it was discovered in the numerical simulations of \cite{MS2001}. The mechanism was further generalised by \cite{MorbCrid2007}, who showed that it requires that: \begin{enumerate}
    \item The interior planet must be the more massive of the pair.
    \item The two planets must share a common gap in their protoplanetary disk so that the inner Lindblad torque exceeds the strength of the outer torque, resulting in a positive net torque on the pair of planets.
    \item The orbital configuration of the pair must be sufficiently compact such that their horseshoe regions overlap. This facilitates the flow of gas from the outer planet's horseshoe region, past the inner planet, and into the inner disk. This maintains sufficient gas at the inner Lindblad resonant positions to sustain outward migration.
\end{enumerate}

The last two points require that the secondary (outer) planet is sufficiently massive and orbits sufficiently close to the primary (inner) planet. Saturn happens to fall within the required mass range and could have orbited sufficiently close to Jupiter when the solar nebula (protoplanetary disk) existed. The most favourable orbital configuration is a 3:2 MMR between the two giant planets, which can produce and maintain outward migration over a long range \citep{Crida2009b}. \cite{Griveaud2023} reviewed the current limitations of the Masset and Snellgrove mechanism in producing outward migration, for example a pair of planets entering into a 2:1 MMR and migrating inward if the mass of the outer planet is too low when it approaches the 2:1 MMR \citep{Dangelo2012} or the relative migration speeds of the two planets are too slow for the outer planet to cross the 2:1 MMR \citep{Chametla2020}. \cite{Griveaud2023} went on to show that if the disk viscosity is lower ($\alpha$\footnote{This is the standard parameterisation of turbulent strength based on \cite{SS73}.} $\sim 10^{-4}$) than is often assumed in previous disk models ($\alpha$ $\sim 10^{-2}-10^{-3}$), the pair of planets are more likely captured into a 5:2 or 2:1 MMR due to their slow relative migration.

Based on this more recent work, it is equally as likely that the grand tack did not happen, as it appears that the Masset and Snellgrove mechanism requires physical properties that do not align with our current understanding of the early Solar System. The purpose of this Letter is not to argue in favour or in opposition of the grand tack model, but to search the exoplanet data for other systems that show possible evidence for being in an orbital alignment indicative of the Masset and Snellgrove mechanism. This was already done by \cite{MorbCrid2007}, but at a time when our knowledge of the full extent of the exoplanet population was limited. I searched through the exoplanet population data to identify systems of (at least two) giant planets, with masses $>0.3$ M$_{\rm Jupiter}$ (approximately Saturn's mass), and report their current orbital period ratio in Sect. 2. In Sect. \ref{sec:discussion} I explore whether any of these systems show the possibility of having supported planets in a Masset and Snellgrove mechanism-driven stage of evolution, and conclude on this short exploration.

\begin{figure*}
    \centering
    \includegraphics[width=0.9\textwidth]{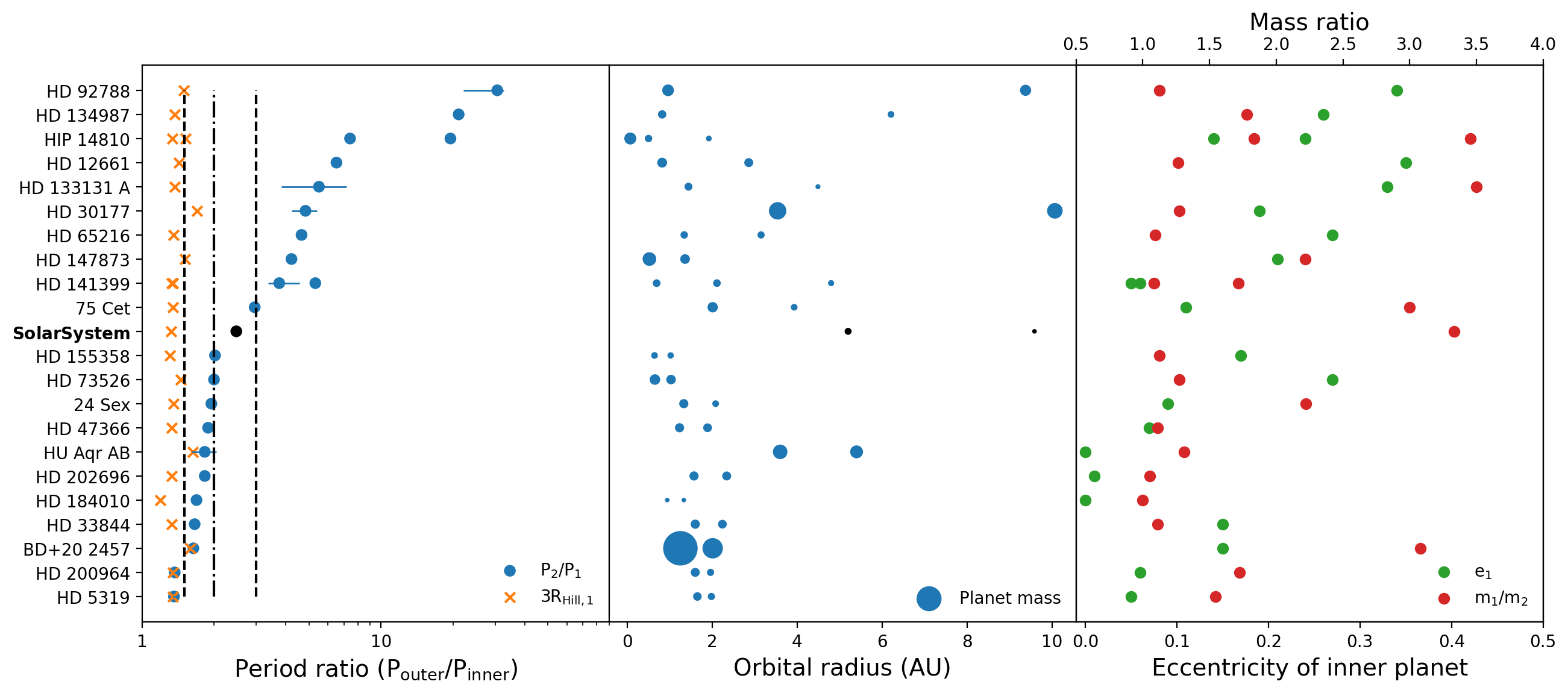}
    \caption{Orbital data extracted from the NASA Exoplanet Archive on May 2, 2024, for a set of planetary systems that (at least) partially meet the requirements for the Masset and Snellgrove mechanism. \textbf{Left:} Period ratio of planet pair in a `grand tack' mass arrangement (i.e. the interior planet is more massive) in the exoplanet population, as well as for Jupiter and Saturn (labelled `Solar System'). Some systems contain more than one pair and are thus represented by multiple points. The orange crosses show the period ratio if the secondary planet orbits at an orbital distance of 3 Hill radii of the primary. Three of the systems (at the bottom of the plot) orbit near this arrangement. The three vertical lines show, from left to right, orbital period ratios of exactly 3:2, 2:1, and 3:1, respectively. \textbf{Middle:} Orbital separation of each planet from their host star in units of AU. The size of each marker is proportional to the mass of the planets; the black points show the orbital structure of Jupiter and Saturn in our own Solar System. \textbf{Right:} Eccentricity of the inner planet as well as the mass ratio of each planetary pair. A wide range of eccentricities is seen for the exoplanet population; however the planets with high period ratios (top of the figure) seem to generally have higher eccentricities on average.}
    \label{fig:period_ratio}
\end{figure*}

\section{Searching the exoplanetary data} \label{sec:exodata}

The exoplanet population data were retrieved from the NASA Exoplanet Archive\footnote{http://exoplanetarchive.ipac.caltech.edu} on May 2, 2024. It contains 933 planetary systems that had both planetary masses and semi-major axes available. The vast majority did not have planetary parameters in line with the Masset and Snellgrove mechanism: either they do not contain two sufficiently massive planets (M$_{\rm planet} > $ M$_{\rm Saturn}$) or the inner-most planet was not the more massive. This left 29 pairs of planets that meet requirement (1). In some planetary systems (such as HD 141399) there are multiple pairs of giant planets. In these cases, I filtered out pairs of planets that clearly would not interact via the Masset and Snellgrove mechanism because they have a third planet orbiting at radii in between them; this left me with a total of 23 pairs of planets. Where multiple entries of orbital period and planetary mass are included, their averaged values and their largest reported uncertainties were used in the following analysis.

The orbital period ratio of each planetary pair is shown in the left panel of Fig. \ref{fig:period_ratio}. On the y-axis the planetary systems are ordered from the bottom up, beginning with the lowest orbital period ratio. Along with the period ratio (blue points), I computed what the period ratio would be if the outer planet were orbiting at a radial distance of 3 Hill radii\footnote{The Hill radius is defined as R$_{\rm Hill} = a_p(M_p/3M_*)^{1/3}$, where $a_p$ and M$_p$ are the mass and semi-major axis of the planet.} larger than the orbital radius of the inner planet. The width of the gap opened in the protoplanetary disk by giant planets has been found in numerical simulations to be between 7 and 10 R$_{\rm Hill}$ \citep{Rosotti2016}, so a distance of 3 R$_{\rm Hill}$ would correspond to planet formation near the outer edge of the gap carved by the inner planet. The middle panel of Fig. \ref{fig:period_ratio} shows the orbital separation from the host star for each of the planets in each system. The right panel shows the mass ratios of each planet pair and the inner planet's eccentricity.

Along with the exoplanet data, I include the period ratio of Jupiter and Saturn. We can see that the Solar System giants have a period ratio that places them near the middle of the rest of the population. There are a number of planetary systems with smaller period ratios, ranging from near 2:1 MMR down to lower than 3:2 MMR. The bottom three systems correspond to a period ratio indicative of an orbital separation of 3 of the inner planet's Hill radii. The rest all seem to have period ratios that fully span the range between 3:2 and 2:1 MMR. Above the Solar System, the exoplanet data seem to be separated by no less than a 3:1 MMR, with many systems showing progressively larger period ratios. The exoplanetary data appear to have a gap in period ratios between 2:1 and 3:1 MMR that our own Solar System fills.

The middle and right panels show more specific orbital and planetary properties for each system. The middle panel shows the orbital semi-major axis for each of the planets in each system. The right panel shows the mass ratio of each pair of planets within each system (in red) and the eccentricity of the inner planet in each pair (in green). Many of the planetary systems show mass ratios that are of the order of a few or less. There is a similar spread in the eccentricity; however, the planetary systems that have higher orbital period ratios also tend to have higher orbital eccentricities. At the lower end of the orbital period ratio, the planetary systems tend to have similar masses, mass ratios, and orbital separations (with the one obvious exception of BD+20 2457).

\section{ Discussion and conclusion }\label{sec:discussion}

\subsection{Orbital period ratios in the exoplanet data}

The Masset and Snellgrove mechanism requires a pair of planets to be closely orbiting; this is typically thought to require a 3:2 MMR, though this depends on the local disk properties \citep[][showed that a 2:1 MMR can also lead to outward migration]{Crida2009b}. This orbital architecture, however, may not have survived the following gigayear of orbital evolution. In our own Solar System there is evidence for a restructuring of the orbital architecture of both the gas and ice giants \citep[the so-called Nice model;][]{Tsiganis2005,Gomes2005,Morby2005}, so just because pairs of planets are not currently in a perfect 3:2 MMR does not mean they were not so in the past. 

Recently, \cite{Charalambous2022} studied a similar but opposite set of exoplanets in a similar way as I have done here. Their population is opposite in the sense that they focused on systems where the inner planet is less massive than the outer planet (m$_1$/m$_2 < 1$). They note that these sorts of systems represent $\sim 95\%$ of the exoplanet population, which is consistent with my findings here. They observe a nearly continuous set of planets in between a 3:2 and 3:1 MMR, particularly when the planets are of low mass. Their high-mass pairs also have a variety of period ratios; however, they find a build-up just exterior to the 3:2 and 2:1 MMRs, which they explain as being caused by a small resonant offset produced during the disk migration phase.

In the (much smaller) population of planets studied here, we can see that there is a continuous coverage of period ratios between 3:2 and 2:1 MMR; however, there are no planet pairs (other than Jupiter and Saturn) between 2:1 and 3:1 MMR. It is unclear why this would be the case. However, in the simulations of \cite{Griveaud2023}, the most common outcome of disk migration was a 2:1 MMR, with a 5:2 MMR only occurring in low viscosity disks when the disk scale height was the lowest ($h_0 = 0.035$, i.e. a very cold disk). This may suggest that protoplanetary disks are, on average, thicker than a scale height ratio of $0.035,$ which is consistent with our understanding of disks \citep[see for example the recent review of][]{Miotello2023}.

In Fig. \ref{fig:period_ratio}, the bottom three planetary systems seem to be consistent with the second, outer planet having formed near the outer edge of the gap carved by the first. The other planetary pairs, with lower orbital period ratios than that of the Solar System, could be consistent with planets having entered into a low-order MMR (either 2:1 or 3:2) during the disk phase but were later gradually perturbed away from MMR through mutual gravitational interactions. This perturbation, however, is likely too slow for giant planets to be responsible for what we see here. \cite{LithwickWu2012} derived the time evolution of the deviation from perfect MMR:\begin{align}
    \Delta \equiv \frac{P_2/P_1}{p/q} -1,
\end{align} where $p > q$ are integers, and the subscripts $1$ and $2$ denote the inner and outer planets, respectively. Its evolution is \citep[see also][]{Dai2024} \begin{align}
    \Delta (t) \simeq 0.006 \left(\frac{Q}{10}\right)^{-1/3}\left(\frac{M_1}{10M_\oplus}\right)^{1/3}\left(\frac{R_1}{2R_\oplus}\right)^{5/3} \left(\frac{M_*}{M_\odot}\right)^{-8/3} \nonumber\\
    \times\left(\frac{P_1}{5 {\rm ~days}}\right)^{-13/9}\left(\frac{t}{5 {\rm ~Gyr}}\right)^{1/3}\left(2\beta + 2\beta^2\right)^{1/3},
    \label{eq:deltat}
\end{align} where $\beta$ involves the ratio of outer to inner planet masses and the ratio of the square root of their semi-major axes. For tides raised by inertial waves, \cite{Lazovik2024} find that the quality factor ($Q$) of giant planets is of the order of $10^4$, which implies for exoplanet systems between 2:1 and 3:2 MMR that their tidal evolution would have to be many orders of magnitude longer than the age of the Universe to deviate as far from either MMR as they are observed to have done today. The primary reason for this is that many of these planets have large orbital periods ($P_1 >> 5$ days in Eq. \ref{eq:deltat}) and are thus less sensitive to tides raised by their host star. Instead, they likely arrived at their current orbital period ratio while the protoplanetary disk was still present, but how the outer planet migrated through the 2:1 MMR remains a mystery\footnote{At least to me.}.

Recent work has shown that tidal effects can be transported across MMR pairs \citep{Cerioni2023,Charalambous2023}, which could increase the extent of their possible impact on planetary systems \citep[this was also argued for Saturn's moons by][]{Allan1969}. It is also possible that some of these systems are being perturbed by a third body that is currently undetected. In such a case, the pair of orbital period ratios would appear to have drifted from perfect MMR but could be in a three-body MMR \citep[i.e. as shown in][]{Charalambous2023}.

Two planetary systems (HD 73526 and HD 155358) currently appear very near to the 2:1 MMR. The planet pair in HD 73526 was analysed by \cite{Pons2024} in the context of long-term stability for planet pairs very near to exact MMR. They find evidence that this particular system is stable over long timescales, suggesting that this period ratio was inherited from their orbital migration within the protoplanetary disk phase. With their orbital period ratios so close to 2:1 MMR and their orbital radii sitting near 1 AU (see the middle panel of Fig. \ref{fig:period_ratio}), it is unlikely that these two systems underwent grand tack-like evolution. 

The planetary systems with larger period ratios than the Solar System's exist with a wide range of orbital radii and generally a higher inner planet eccentricity than the systems with lower period ratios. The higher eccentricity may suggest that some of these systems (e.g. HD 92788, HD 134987, and HD 330177) have undergone a stage of gravitational scattering indicative of what is known as high eccentricity migration \citep[e.g.][]{WuMurray2003,FabryckiTremain2007,Petrovich2015}. The inner planets have migrated inward due to the gravitational perturbation of their planetary siblings and are currently in the process of circularising. 

Some of these large period ratio systems contain a hot giant planet and a relatively compact system of planets. The period ratios, however, are so large (compared to the 3 Hill radius ratio) that it is unlikely these planets formed in conjunction. Instead, they may have formed in isolation and migrated to their current positions without ever interacting gravitationally. The HIP 14810 system in particular shows an orderly planetary system similar to our own outer Solar System, with a decreasing planetary mass with orbital radius but at smaller orbital periods. It is possible that each planet migrated to its current position without much mutual interaction, making it a true failed grand tack. This would imply that the second planet in the chain either never attained enough mass or never migrated close enough to the inner, more massive, planet to activate the tack.

\subsection{The PDS 70 system}

A popular (in some circles) proto-planetary system is PDS 70, with its two confirmed protoplanets \citep{Keppler2018,Haffert2019} and extended dust \citep{Isella2019,Benisty2021} and gas disk \citep{Facchini2021}. PDS 70b and PDS 70c certainly fit the requirement of having an overlapping disk gap as their mutual gap is nearly devoid of dust \citep{Benisty2021} and depleted in gas emission \citep{Facchini2021}. Their masses are not well constrained; however, their orbital period ratio was found to be near the 2:1 MMR \citep{Wang2021}, which suggests that they similarly did not undergo a grand tack.

Growth models of these planets \citep{Cridland2023} via gas accretion suggest that both planets likely originated from similar positions in the protoplanetary disk and would thus require a sequential formation scenario \citep[as recently modelled by][]{Lau2024}. In this scenario, the inner planet forms first and provides the necessary substructure to seed the formation of the second planet. After the initial formation, the two planets would then migrate to their current position and be captured in the 2:1 MMR.

\subsection{Conclusion}

In this Letter I have explored the current exoplanetary data in search of a system that one could recognise as having evolved through the Masset and Snellgrove mechanism, often called the grand tack model in regards to our Solar System. Solid evidence of this would be a pair of giant planets of at least Saturn's mass orbiting in a 3:2 MMR, with the inner planet being more massive. Such a system does not exist in the available data, and for systems near the 3:2 MMR it would take too long for them to have evolved to their current period ratios if they started at 3:2 when their protoplanetary disk evaporated. 

On the other hand, I do find three systems that have period ratios that are consistent with the outer planet having formed near the edge of the gap carved by the inner planet. This could be evidence for sequential planet formation, as discussed by \cite{Lau2024}. I find a few systems with period ratios between the 3:2 and 2:1 MMRs that are too far from MMR to have evolved there within the age of the Universe but could have migrated to their current orbital period during the disk phase (having passed the 2:1 MMR), or have ended up in this orbital architecture due to gravitational interactions with a third (unknown) planet\footnote{This would be similar in principle to the Nice model.}. There are a pair of planets that appear nearly exactly at the 2:1 MMR and for which theoretical work suggests that their orbital history is stable there.

A final population of exoplanetary systems have large orbital period ratios and tend to have larger eccentricities than the low orbital period systems. Some of these planetary systems may have undergone a stage of evolution indicative of high eccentricity migration, while others are more compact and could have evolved through diverging orbital migration.

There are no known planetary systems, other than our Solar System, that have orbital period ratios between 2:1 and 3:1 MMR. This is likely due to the limited population of exoplanetary systems that were the focus of this Letter and not due to some underlying physical process. Nevertheless, the orbital separations and mass ratios of the Solar System's most massive giants appear to be similar to the rest of the planetary systems studied here.

Similar to Type-I and Type-II migration, the Masset and Snellgrove mechanism results from general gravitational interactions of one or more young planet(s) in their natal protoplanetary disk. As such, we should expect (as we do for Type-I/II migration) that the mechanism will show itself in the known exoplanet population. It is possible, however, that the current population of known exoplanetary systems is biased against systems that underwent the grand tack because the planets in these systems (by definition) orbit at radii similar to or larger than that of our own giant planets. This means that they exist in a range of orbital parameters outside of what is currently detectable by modern-day observatories.

\begin{acknowledgements}

 Thank you to Dr. Charalambous for their thoughtful review of the submitted draft which improved its clarity. Thanks to Jan Harre and Konstantin Batygin whose discussions helped in shaping the final form of this letter. Planet formation and astrochemical studies at the LMU are done as part of the Theoretical Astrophysics of Extrasolar Planets research chair. This project made use of the following software: Astropy \citep{astropy}, SciPy \citep{Scipy}, NumPy \citep{numpy}, and Matplotlib \citep{matplotlib}. 

\end{acknowledgements}

\hfill 

\bibliographystyle{aa} 
\bibliography{mybib.bib} 

\end{document}